# Spin orbit torque switching of antiferromagnet through the Néel reorientation in rare-earth ferrite

T. H. Kim[1,**], S. Hwang[1], S. Y. Hamh[2], S. H. Yoon[3], S. H. Han[4] and B. K. Cho[1*]

[1]*School of Materials Science and Engineering, Gwangju Institute of Science and Technology (GIST), Gwangju 61005, Republic of Korea*

[2]*Center for Bio & Basic Science R&D Coordination, Korea Institute of S&T Evaluation and Planning (KISTEP), Eumseong-gun 27740, Republic of Korea*

[3]*Smart Energy & Nano Photonics Group, Korea Institute of Industrial Technology, 6, Choemdan-gwagiro 208-gil, Bukgu, Gwangju, 61012, Republic of Korea*

[4]*Division of Navigation Science, Mokpo Maritime National University, Mokpo 58628, Republic of Korea*

Correspondence and requests for materials should be addressed to B. K. Cho (*chobk@gist.ac.kr).

We suggest coherent switching of canted antiferromagnetic (AFM) spins using spin-orbit torque (SOT) in small magnet. The magnetic system of orthoferrite features biaxial easy anisotropy and the Dzyaloshinskii Moriya interaction, which is perpendicular to the easy axes and therefore creates weak magnetization ($m$). A damping-like component of the SOT induces Néel reorientation along one of the easy axes and then exerts torque on $m$, leading to tilting of the Néel order $l$. The torque on the magnetization becomes stronger due to coupling with the induced Oersted field or the field-like component of the SOT, enhancing the tilting of $l$. Therefore, $l$ is found to experience deterministic switching after the SOT is turned off. Based upon both numerical and analytical analysis of the coherent switching, XOR logic gates are also found to be implemented in a single magnetic layer. In addition, we investigate how magnetic parameters affect the critical reorientation angle and current density in a simple layered structure of platinum and a canted AFM. Our findings are expected to provide an alternative spin-switching mechanism for ultrafast applications such as spin logic and electronic devices.

**Present address: Department of Electrical and Computer Engineering, National University of Singapore, 117576 Singapore

# I. INTRODUCTION

Electrical switching of antiferromagnets (AFMs) or antiferromagnetic insulators (AFIs) is of great interest for ultrafast memory device applications with potential write times approaching sub-nanosecond times or a few nanoseconds [1-6]. This is reflected by the fact that the terahertz dynamics of the AFM are dominated by a strong exchange field, which is several orders of magnitude larger than that of ferromagnetic (FM) materials.

The issues of AFM spintronic memory devices are twofold [5]: (i) reliable electrical reading of the magnetic state of the AFM (e.g., the Néel order parameter) and (ii) a robust method to electrically modify the AFM state. This work focuses on the second issue by proposing a robust switching method in the AFI. Because of the lack of conduction electrons in the AFI, the conventional spin transfer torque where currents are directly injected into magnetic materials is not feasible. Instead, spin orbit torque (SOT), which can be split into two generic classes, namely, damping-like and field-like SOT, is a possible way to transfer spin momentum carriers into the AFI [7, 8, 9].

In the FM, one of SOT-driven switching schemes exploits the anti-damping effect $\sim \mathbf{m} \times (\mathbf{m} \times \mathbf{p})$, where spin polarization ($\mathbf{p}$) is collinear with the magnetic easy axis. After reversal, the role of SOT is counter anti-damping, making the reversed state fixed [9, 10, 11]. This scheme, however, is not the case in the AFM because the anti-damping effect acts on an opposite AFM sublattice, leading to auto-oscillation rather than switching [12].

Another scheme is proposed where FM and AFM easy axis are orthogonal to $\mathbf{p}$. Therefore, SOT does not directly compete with the damping torque and can be understood as an equivalent magnetic field, $\sim \mathbf{m} \times \mathbf{p}$. In incoherent switching, the FM requires a magnetic field parallel to current direction to induce domain wall expansion for switching [13-16] and in coherent switching, the FM needs small perturbation such as the magnetic field for switching [17, 18]. Unfortunately, the AFM shows different dynamics, at least in the coherent picture; at the critical current density, a steady oscillation, similar to an auto-oscillation, occurs because the AFM has inertia [5, 6, 7].

In these oscillatory behaviors of the AFM, switching is realized with a properly controlled current pulse; an appropriate duration to release the magnetization at the desired final state should be estimated [4, 21-27]. It requires (i) strong coherency of the magnetic moments with respect to thermal agitations [28] in the device and (ii) a broad tolerance of current-pulse duration and strength. In particular, it has narrow tolerance in a low damping system such as the AFI, and the desired final state is not easy to determine [21].

In this work, we propose a robust $\pi$ switching scenario of the canted antiferromagnets (CAFs) through Néel reorientation, which is triggered by SOT. The target material is an yttrium ferrite (YFeO$_3$). The magnetic system of YFeO$_3$ is of biaxial anisotropies (e.g., the *x* and *z* directions) and a Dzyaloshinskii Moriya interaction (DMi). The superexchange interaction occurs between two $Fe^{3+}$ ions, which are separated by an $O^{2-}$ ion. Below the Néel temperature $T_N$=645 K [30], the Fe spins order antiferromagnetically with an easy axis along the *x* direction of the crystal, with a small canting (~0.4 degrees) along the orthorhombic *z* direction where weak ferromagnetism is caused by three mechanisms: the exchange interaction, the DMi and single-ion crystalline anisotropy.

As shown in Fig. 1(a), the single crystal YFeO$_3$ is sandwiched by two layers of platinum (Pt) and SrRuO$_3$/SrTiO$_3$ (001) substrate where it results in the (010) direction of YFeO$_3$ and the electric field control of DMi of YFeO$_3$ [29]. Our device is a small single-domain nanomagnet (typically, 50×50 nm$^2$ or less) where the domain wall formation is energetically unstable. To clearly demonstrate the switching scenario, we focus on a damping-like SOT and discuss the effect of the field-like component of the SOT.

The Néel reorientation is taken into account energetically. For example, SOT increases the exchange, DMi and anisotropy energy of a CAF system. As a result, the Néel phase is shifted by $\varphi$ (see Fig. 1(b)) because of the momentum conservation principle. In $\varphi=\varphi_{RE}$, the magnetic easy axis (i.e., *x* direction) is energetically unstable, the anisotropy along the *z* direction becomes the new easy axis, and *l* is reoriented along the *z*-direction (see Fig. 1(c)). We note that an additional torque between *m* and the damping-like component of the SOT breaks symmetry, giving rise to tilting of *l* and *m*. In addition, the current-induced Oersted field along the *z*-axis breaks the symmetry. After the SOT is turned off, *l* is robustly reversed (see Fig. 1(d)). Because DMi is controlled by an electric field [29, 31, 32], the SOT and electric field can be combined to implement an XOR gate in a single AFM layer without an additional magnetic layer or a magnetic field.

## II. MODEL OF ANTIFERROMAGNETIC DYNAMICS

This section presents the magnetization switching of YFeO$_3$. YFeO$_3$ is a canted antiferromagnet, two spins of which are antiparallel to each other and are slightly canted along the *z-axis*. Their energetic description is given as

$$U = -J\mathbf{s}_1 \cdot \mathbf{s}_2 + \mathbf{D}_y \cdot (\mathbf{s}_1 \times \mathbf{s}_2) + K_x[(s_{1,x})^2 + (s_{2,x})^2] + K_z[(s_{1,z})^2 + (s_{2,z})^2] - gu_B \mathbf{h} \cdot (\mathbf{s}_1 + \mathbf{s}_2). \quad (1)$$

where $\mathbf{s}_i = \mathbf{S}_i/S_0$ ($\mathbf{s}_2 = \mathbf{S}_2/S_0$) is normalized by the magnitude of the magnetization $S_0 = |\mathbf{S}_1|(=|\mathbf{S}_2|)$, and $\hbar$ is the reduced Plank constant. From the left side, $J$ is the exchange energy, the sign of which is negative, $\mathbf{D}_y$ is the DM vector along the *y-axis*, $K_x$ ($K_z$) is the uniaxial anisotropy constant with easy axis $x$ ($z$) axes, and $\mathbf{h}$ is the Oersted field induced by the current flow and forms Zeeman energy by coupling with Landé-$g$ factor, $g$ (=2), and Bohr magneton, $u_B$. By defining magnetization $\mathbf{m} = (\mathbf{s}_1 + \mathbf{s}_2)/2$ and the Néel order $\mathbf{l} = (\mathbf{s}_1 - \mathbf{s}_2)/2$ and converting energy into frequency units divided by reduced Plank constant $\hbar$ (e.g. $\omega_E = J/\hbar$), the coupled LLG equations based on two sublattices are described as [21, 33]

$$\dot{\mathbf{m}} = [\omega_{D,y}\hat{\mathbf{y}} \times (\mathbf{l} \times \mathbf{m}) + \omega_{K,x}\hat{\mathbf{x}} \times (m_x\mathbf{m} + l_x\mathbf{l})] + \omega_{K,z}\hat{\mathbf{z}} \times (m_z\mathbf{l} + l_z\mathbf{m})] + ...$$
$$\beta(\mathbf{l} \times \dot{\mathbf{l}} + \mathbf{m} \times \dot{\mathbf{m}}) + \omega_s(\mathbf{m} \times (\mathbf{m} \times \mathbf{p}) + \mathbf{l} \times (\mathbf{l} \times \mathbf{p})) + \omega_H \mathbf{m} \times \mathbf{h} \quad (2a)$$

$$\dot{\mathbf{l}} = [2\omega_E(\mathbf{l} \times \mathbf{m}) + \omega_D(\mathbf{l} \times (\mathbf{l} \times \hat{\mathbf{y}}) - \mathbf{m} \times (\mathbf{m} \times \hat{\mathbf{y}})) + \omega_{K,x}\hat{\mathbf{x}} \times (m_x\mathbf{l} + l_x\mathbf{m}) + ...$$
$$\omega_{K,z}\hat{\mathbf{z}} \times (m_z\mathbf{l} + l_z\mathbf{m}) + \beta(\mathbf{m} \times \dot{\mathbf{l}} + \mathbf{l} \times \dot{\mathbf{m}}) + \omega_s(\mathbf{m} \times (\mathbf{l} \times \mathbf{p}) + \mathbf{l} \times (\mathbf{m} \times \mathbf{p})) + \omega_H \mathbf{l} \times \mathbf{h} \quad (2b)$$

Here, $\beta$ is the phenomenological damping constant, and $\omega_s(=J_c\sigma)$ is the spin-orbit torque strength proportional to the product of the current density $J_c$ (A/m$^2$) and the proportionality factor σ. In simple two layers, σ is defined as $\frac{\gamma\hbar\theta_H}{2et_zM_s}$ [34], where $M_s$ is the saturation magnetization (A/m), $t_z$ is the thickness (m) of the YFeO$_3$, $\theta_H$ is the effective spin Hall angle, $\mathbf{p}$ is the spin current polarization $\mathbf{p}=(p_x, p_y, p_z)$ and $\mathbf{h}$ is the oersted magnetic field $\mathbf{h}=(h_x, h_y, h_z)$ with a field strength $\omega_H$. In an exchange limit ($|\omega_E| > \omega_D \gg \omega_{K,x}, \omega_{K,z}$ and $\omega_H$), taking the cross product of $\mathbf{l}$ to Eq. 2b and neglecting the high-order elements, we obtain the general analytical relations between $\mathbf{m}$ and $\mathbf{l}$:

$$\frac{\dot{\mathbf{l}} \times \mathbf{l}}{2\omega_E} \sim \{l_z\dot{l}_y - l_y\dot{l}_z - \omega_D l_z, -l_z\dot{l}_x + l_x\dot{l}_z, l_y\dot{l}_x - l_x\dot{l}_y + \omega_D l_x + \omega_H)/(2\omega_E)\} = \mathbf{m} \quad (3)$$

At equilibrium, $\mathbf{m}$ is aligned along the *z-axis*, the magnitude of which is $m_z = \omega_D l_x/(2\omega_E) = D_y l_x/(2J)$. We express the Néel order parameter as $l = (\sin\theta\cos\varphi, \sin\theta\sin\varphi, \cos\theta)$. By injecting a spin current into the AFM, a pendulum-like equation of motion is selectively driven depending on the spin current polarization [21]; for example, when *p* or *H* is parallel to the *z* (*y*) axis, the qAFM (qFM) mode oscillating on the *xy* (*xz*) plane with dynamic parameters such as the azimuthal angle $\varphi(t)$ and the polar angle $\theta = \pi/2$ ($\theta(t)$ and $\varphi = 0$) is excited. However, when the reorientation is considered in the qAFM mode, we can obtain the combined processional mode of *l* on a 3-dimensional plane with dynamic parameters $\varphi(t)$ and $\theta(t)$:

$$\ddot{\theta} + 2\beta\omega_E\dot{\theta} - \sin[2\theta](\omega_{qFM}^2 - \omega_{qAFM}^2 \sin[\varphi]^2)/2 = 0 \tag{4}$$

$$\ddot{\varphi} + 2\beta\omega_E\dot{\varphi} + \sin[2\varphi]\omega_{qAFM}^2/2 + \omega_D\omega_{H,z}\sin[\varphi] = 2\omega_E\omega_{S,z} + \dot{\omega}_{H,z} \tag{5}$$

where $\omega_{qAFM}^2 = 2\omega_E\omega_{K,x} + \omega_D^2$ and $\omega_{qFM}^2 = 2\omega_E(\omega_{K,x} - \omega_{K,z})$. In the application of the static spin current and the Oersted field, the final phases of *l* for qAFM mode are obtained as

$$\varphi \sim \frac{1}{2}\arcsin\left(\frac{4\omega_s\omega_E}{2\omega_E\omega_{K,x} + \omega_D^2}\right) \tag{6}$$

When $\varphi = \pi/4$ corresponds to $\omega_s = \omega_{qAFM}^2/2$, the AFM undergoes a steady oscillation where a precession around the x-axis turns into a precession around the z-axis in [33]. The sign of $\sin[2\theta]$ in Eq. (4) is initially negative, implying that $\theta(0\text{ ps}) = \pi/2$ and is affected by the qAFM mode in Eq. (5). For $\omega_{qFM}^2 < \omega_{qAFM}^2 \sin[\varphi]^2$, the Néel polar angle evolves from $\pi/2$ to $0$. Here, $\varphi_{RE}$ is obtained as $\omega_{qFM}^2 = \omega_{qAFM}^2 \sin[\varphi_{RE}]^2$

$$\varphi_{RE} = \arcsin\left(\frac{\omega_{qFM}}{\omega_{qAFM}}\right). \tag{7}$$

From Eq. 6 and Eq. 7, the critical current density for reorientation is obtained as

$$J_{RE} = \omega_S/\sigma = \frac{\omega_{qAFM}^2 \sin\left(2\arcsin\left(\frac{\omega_{qFM}}{\omega_{qAFM}}\right)\right)}{4\omega_E\sigma\hbar}. \tag{8}$$

Here, $\theta = 0$ corresponds to the case where *l*(*m*) is relaxed along the ***x-axis*** (***z-axis***), and $\theta = \frac{\pi}{2}$ describes *l*(*m*) along the ***z-axis*** (***x-axis***). For the proper parameters, $J_{RE}$ can be much smaller than $J_{st}$, so reorientation occurs before steady oscillation. It can be utilized for deterministic switching triggered by SOT.

## III. RESULTS AND DISCUSSION

In all simulations, $K_x$ is larger than $K_z$. At equilibrium, the Néel order *l* is aligned along the *x-axis*, while *m* is formed along the *z-axis* due to DMi. When the current flows along the *x* direction, the spin current is injected into the top layer with polarization $p_z$ (see Fig. 1(a)), and the Oersted field induced by Ampere's Law with $H_{Oe}=\mu_0 J_c t_{Pt}/2$ is applied along the *z-axis*. Both external stimuli excite the qAFM mode; therefore, *l* is relaxed on the *xy* plane with a finite $\varphi$, as illustrated in Fig. 1(b).

We first investigate the reorientation of *l* on the *xy* plane to the *z-axis* in the application of a continuous current density and an Oersted field. In YFeO3, $\omega_E$ =97 THz ($J = 63.7$ meV), $\omega_D$ =2.4 THz ($|D_y|$ = 1.6 meV), $\omega_{K_x}$ =33 GHz ($K_x = 22$ $\mu$eV) and $\omega_{K_z}$ =15 GHz ($K_x = 9.9$ $\mu$eV) in units of $2\pi$ [29, 34]. With parameters taken from YFeO3, $\varphi_{RE}$ is calculated as 34.8 degrees. The critical current density to reorient *l* and the estimated Oersted field are calculated as approximately $J_c = 1.99\times10^{12}$ Am$^{-2}$ and $H_{Oe}$=63 Oe or $\omega_H = 1.1$ GHz in units of $2\pi$ where $t_{Pt}$ = 5 nm, $g = 2$, $\theta_H = 0.2$ for polycrystalline Pt [34] and $t_z$=2 nm, $M_s$=397 kA/m for YFeO3 [30].

Figure 2 shows that AFM switching is triggered by SOT with *p*//-*z* when $D_y >0$ and $l_x>0$; thus, $m_z>0$. The dynamic variable is $\varphi(t)$ until $\varphi(t=58$ ps$)=\varphi_{RE}$ (see Fig. 2(a)). However, since *t*=58 ps, *l* starts to be reoriented along the *z-axis* ($\theta$~0). This period from 15 ps to 58 ps is defined as the incubation time $\tau_{in}$ when the *z* component of the Néel vector drops by 3% from the equilibrium phase. Similarly, *m*, initially aligned along the *z-axis*, is realigned to the *x-axis*, as shown in Fig. 2(b). After the pulse is turned off, *l* is deterministically switched to the opposite direction due to the tilting component of *l*, $l_x$~-0.026(see the inset of Figure 2(a)); all trajectories on the unit sphere are described with respect to *l* and *m* in Fig. 2(c) and $s_1$ and $s_2$ in Fig. 2(d).

Figure 2(e) is an energetic description of reorientation. The reorientation mechanism is characterized by energetic competition among the exchange, DM interaction, anisotropy and spin orbit torque. The normalized magnetic energies are introduced: the exchange gain, $\Delta E_{Ex} = \left(J(s_1(t)\cdot s_2(t) - s_1(0)\cdot s_2(0)) + D_y \cdot (s_1(t)\times s_2(t) - s_1(0)\times s_2(0))\right)/K_x$, the anisotropy, $\Delta E_K = \Delta E_{K,x} + \Delta E_{K,z} = \left(-K_x(s_{1,x}(t)^2 + s_{2,x}(t)^2) - K_z(s_{1,z}(t)^2 + s_{2,z}(t)^2)\right)/K_x$ and total energy $\Delta E_T = \Delta E_{Ex} + \Delta E_K$. To clearly describe the reorientation, we set two arbitrary time points, $t_2$ and $t_3$, which are assumed to complete decay and reorientation at $t=t_2$ and $t=t_3$ (e.g., $t_2$ =45 ps and $t_3$=100 ps), respectively. When SOT with $\pm p_z$ excites the qAFM mode, it induces spin canting or $\pm\delta m_z$ temporally. Then, the Néel phase is shifted by $\varphi(t_2)$=arctan($\pm\Delta l_y/l_x$) or rotated around the **z-axis** (see Fig. 1(b) and Fig. 2(a)) because of momentum conservation; the exchange energy gain from SOT is transferred partially to increase the anisotropy energy (see Fig. 2(e)). Here, $|m_z(t_1)|$ <$|m_z(0 ps)|$ because $l_x$=1-$|\Delta l_y|$ and $m_z$=$l_x D_x/(2J)$. The finite $\Delta\varphi$ and $\Delta m_z$ correspond to $\Delta E_K$ and $\Delta E_{Ex}$, respectively. Before (after) reorientation, the magnetic easy-axis is the *x-axis* (z-axis), which means that $\Delta E_{K,x}$ ($\Delta E_{K,z}$) represents $\Delta E_K$. Note that, without the contribution of $\Delta E_{Ex}(t=t_2)$, $E_K(t=t_2) < E_K(t=t_3)$ so that the reorientation is not allowed, which means that the exchange gain plays a crucial role in the reorientation. In a collinear AFM, however, reorientation takes place only in the case that $\Delta E_K(t=t_2) > \Delta E_K(t=t_3)$. The exchange gain induced by SOT is fully transferred to $\Delta E_K$ because $|\omega_E| >> \omega_{K,x}$ or collinearity is strongly secured. Thus, there is no contribution of the exchange gain (see APPENDIX A for details). From the two AFM systems, we can conclude that the DMi, the energy of which is not much smaller than the exchange energy ($|\omega_E| > \omega_D >> \omega_{K,x}$), prevents the (i) AFM state from being collinear and therefore transfers momentum that is partially injected by SOT to anisotropy.

After reorientation, it is found numerically that a damping SOT with $p_z$ exerts additional torque on $m_x$, leading to tilting toward the *z* direction. When a SOT with $p_z$ induces a temporal $m_z$, the exchange energy gain by $m_z$ is transferred partially to the anisotropy energy; *l* is rotated around the hard axis (e.a. **x-axis**), so that the $l_y$ component is induced (see the inset of Fig. 2(a)). Therefore, at $t=t_3$, *l* (*m*) is slightly tilted along the **x-axis** and **y-axis** (**z-axis**), as shown in the inset of Fig. 2(a) (Fig. 2(b)). As expressed in Eq. 3, *m* is coupled with the magnetic field. Therefore, the Oersted field $H_z$ induces an additional $m_z$ from $m_x$; therefore, *l* is strongly biased to the **x-axis**. Figure 2(f) shows the relationship between the magnetic field and the induced $l_x$. A field-like component of SOT with $p_z$ plays the same role as the Oersted field. For example, when the field-like component are $0.3\omega_s$, $|l_x(t=t_3)|$=0.35 and $|m_z(t=t_3)|$=0.004, which is more deterministic. This is not the case in a collinear AFM without coupling between SOT and weak

magnetization. In addition, the magnetic field along the *z-axis* does not induce magnetization along the *z-axis* in the reoriented state, except for the spin flip by a strong magnetic field, indicating that DMi provides an efficient way to break symmetry.

In the reoriented state, a biased $l_x$ depends on ***p***(//***H*z**) and ***D***. For example, for the (±) sign of $D_y$ and the ± sign of $p_z$ result in $(\pm) \pm \Delta l_x$. Therefore, by controlling the sign of the $D_y$ and $p_z$ vectors, we can easily implement the XOR gate on a single magnetic layer, as shown in Figure 3. It should be emphasized that, for all cases, the sign of the reoriented $l_z$ does not rely on the sign of $D_y$ or $p_z$ (see Figure 3) and is affected by the final phase of ***l*** just before reorientation. For example, in the various damping constants, the sign of $l_z$ is randomly chosen. However, $m_z$ is induced by SOT (with $p_z$). Therefore, the sign of $l_x$ is determined by the signs of $D_y$ and $p_z$. Therefore, the switching process follows the characteristics of a diode; regardless of the initial state of $l_x$ ($l_x$>0 or $l_x$<0), when $p_z$>0, the resulting $l_x$ is positive.

The DM vector can be controlled by the electric field in multiferroic $YFeO_3$, which is reported in [29, 31, 32]. In fact, only polycrystalline $YFeO_3$ is reported to have multiferroicity. However, in single crystal $SmFeO_3$[36], reverse DM interaction is coupled with the electric polarization ***P*** [36] where four sublattices or two spin pairs have net ***P*** along ***y-axis*** or $\boldsymbol{P}_1 + \boldsymbol{P}_2 \sim \hat{\boldsymbol{e}}_{14} \times (s_1 \times s_4) + \hat{\boldsymbol{e}}_{23} \times (s_2 \times s_3) \sim \hat{\boldsymbol{y}}$ and is perpendicular to ***m*** and ***l***. Therefore, the electric field can be applied along ***y-axis*** and electric-field control of the DM vector will be realized in multiferroic $YFeO_3$ with single crystallinity.

We investigate parameter dependences on $D_y$ and $K_z$ to examine how $J_{RE}$ and $\varphi_{RE}$ vary. The results are shown in Figure 4. For a constant anisotropy of $K_z=K_{z0}$ in Fig. 4(a), a high $D_y$ contributes to the exchange gain more than the anisotropy when CAF is excited by SOT; therefore, reorientation takes place at a low $\varphi_{RE}$. However, a high $D_y$ requires a high SOT strength to reach $\varphi=\varphi_{RE}$ due to the strong potential barrier $\omega_{qAFM}/2$ in qAFM mode, as expressed in Eq. (4). For a constant DM energy of $D_y=D_{y0}$ in Fig. 4(b), an increase in $K_z$ contributes to a decrease in $\Delta E_{K,z}(t=t_3)$ or the potential barrier for reorientation so that Néel reorientation occurs in a low $\varphi_{RE}$ and $J_{RE}$. Proportionality $n_1$ ($n_2$) indicates the ratio of variables $D_y$ ($K_z$) and $D_{y0}$ ($K_{z0}$), where $n_1$=0.382 ($n_2$=0.349) indicates the reorientation limits; for $n_1$<0.382 ($n_2$<0.349), $\varphi_{RE}$>$\pi/4$ and a steady oscillation occurs rather than a reorientation. For $n_2$=2.22, the magnetic anisotropy becomes an easy plane on the *x-z* plane because $K_{x0}$=2.22$K_{z0}$.

As mentioned above, there is no energetic preference for the reorienting direction. However, incubation time is affected by damping constant and external stimuli. We performed simulations for various damping constants and current densities. Figure 4(c) shows that, at a constant $\beta$, $\tau_{in}$ decreases exponentially against $J_c$, indicating that high $J_c$ increases $\Delta E_T(t=t_2)$ and promotes reorientation. In principle, when $\Delta E_T(t=t_2) = \Delta E_T(t=t_3)$, $\tau_{in}$ will be infinite. On the other hand, at a constant $J_c$, a high $\beta$ increases $\tau_{in}$, meaning that high energy loss delays reorientation. In this paper, the Oersted field $H_z$ effect is excluded because $H_z$ changes $\tau_{in}$ by ~1 % (not shown).

It should be emphasized that an additional torque effect on *m* is not described in the analytical results in Eq. 4 because of neglecting the high-order elements related to DMi (see Figs. 5), whereas in the collinear system, the numerical and analytical results are perfectly identical to each other (see Figs 6 in APPENDIX. A for details). However, the analytical results of equations 4 and 5 provide the information about $J_{RE}$ and $\varphi_{RE}$.

To trigger the reorientation of *l*, the current shape is carefully considered since the AF has the characteristics of oscillators with inertia [6, 7]. For example, when $J_{c,Re}$ is applied abruptly, it skips a reorientation mode and moves to a steady oscillation mode because the initial phase with $\varphi=0$ degrees is always phase-matched with external stimuli, so *l* has strong inertia and reaches 45 degrees or more. However, the realistic shape has a finite rise time, and the CAF undergoes phase mismatching to the SOT pulse, leading to a reorientation rather than a steady oscillation. In this calculation, the pulse rise time is approximately 5 ps, which is two or three times longer than one period in qAFM mode: $1/\omega_{qAFM}$ ~1.7 ps. A rise time of more than 5 ps is preferred, and the fall time does not affect the switching scenario. Our strategy for triggering the reorientation takes advantage of the change of the dynamic plane from the *xy* plane to the *xz* plane. One might doubt the possibility of reorienting *l* on a single plane; for example, in qFM mode, where the precession plane is on the *xz* plane with an order parameter $\theta(t)$, steady oscillation occurs at $\theta = \pi/4$ prior to reorientation at $\theta = \pi/2$.

It is necessary to discuss the switching current density and the electrical detection of AFM state. The critical current density in an incoherent switching scheme using Pt is known to be ~$5\times10^{11}$ A/m$^2$ [8]. However, using the noble metal with high charge-to-spin conversion efficiencies, the switching current density can be reduced to ~$6\times10^9$ A/m$^2$ [37]. Even though our work is performed in coherent regime, we believe the switching current density would be reduced further using noble metal. Regarding to the detection of an uniaxial AFM state, it is difficult to adopt the spin Hall resistance (SMR) because SMR requires at least two easy axes to distinguish the magnetic states. For example, a-$Fe_2O_3$/Pt [36], NiO/Pt on MgO [37] or on STO [38] and CoO/Pt on MgO (001) [39] can measure the SMR to detect the AFM states by adopting the intrinsic triaxial anisotropies and the strain-induced biaxial anisotropy. It should be noted that the SMR has a drawback, which is not clearly detect the magnetization state due to the resistive switching [36] when a huge current is injected from Pt. For this, it is suggested to anneal the sample using a high current or deposit the magnetic film with Pt grown *in situ* to remove an artifact by Pt [39, 40]. Even though $YFeO_3$ is known to have the two easy axes but effectively, one dominant easy axis works along ***x-axis***, the magnitude of which is three times larger than that along ***z-axis*** [30]. It favors the two-fold magnetic anisotropy of the $YFeO_3$ layer. Here, it can be considered to introduce the external strain effect to make the two different easy axes for SMR measurement [39]. However, in this study, we focused on the intrinsic magnetic property. The recent researches give a hint; the spin pumping of qFM mode (0.3 THz) shows different polarities in voltage which depends on the weak magnetization and thereby the Néel state [43, 44].

Finally, it remains to compare the performance of our proposed device with other schemes regarding the switching mechanism. Basically, our proposed mechanism is realized in the coherent process; the final Néel phase of our device is determined depending on both the DM vector, controlled by the electric field and SOT polarization. However, in most cases the magnetization switching was demonstrated by an incoherent switching process. In the NiO/Pt and Pt/NiO/Pt [1, 4, 45], the domain-wall-mediated switching lowers the switching current density. They exploit the anti-damping SOT regardless of the crystal symmetries [4] or Néel SOT that requires the system with global centro-symmetry and broken sublattice inversion symmetry [1]. Even though the coherent switching has drawbacks such as the high switching current density, the switching speed should be considered. Basically, the domain wall motion is the motion of magnetic arrays when their precession is over. Its typical speed is estimated by the domain wall velocity. A recent and fastest domain wall speed is reported in 5.7 km/s in the ferrimagnet CoGd [8]. Our device switching time is inversely proportional to the resonant frequency (~THz), leading to tens of picoseconds.

As another mechanism, the voltage controlled switching scheme is well-known as an energy efficient way to control the magnetization. Recently, they have exploited the magneto-electric effect [46] in $Cr_2O_3$/Pt, which is similar to our proposal, or the anisotropy modified by the electric field in B-doped $Cr_2O_3$ [47] and $TmFeO_3$ [48]. However, B-doped $Cr_2O_3$ has still a low Néel temperature ($T_N$~400 K) and the switching of $TmFeO_3$ by the electric field of the femtosecond laser, the electric field strength of which is amplified by the designed antenna pattern on the bulk crystal at the Morin transition temperature, is far from the industrial level. To be a viable device, our SOT based device with high $T_N$, which is added by the electric field means, is more promising because of the implementation of a XOR gate in the simple structure. Although our switching mechanism has a drawback regarding to the sample size (e. a. layer area < 60 $nm^2$), which is smaller than industrial level, we believe that the energetic reorientation of the Néel order would work in the device size that corresponds to the incoherent picture and be an interesting topic for future research, but is not the focus of the present work.

## IV. CONCLUSION

In this work, we investigated the switching of canted antiferromagnets through spin-orbit torque in a canted antiferromagnetic system. The damping-like SOT and DMi are found to play a key role in both the reorientation of *l* and the symmetry break for deterministic switching. For example, the initial easy axis becomes unstable when the critical current density is applied in the qAFM mode so that *l* changes its orientation along the *z-axis* or the new easy-axis. then, *l* feels the deterministic switching with the help of additional torque between the SOT and weak magnetization after SOT is turned off. This tilting component is programmable by tuning the spin current polarization and DMi vector. Once the DMi can be controlled by an external electric field, a combination of current and electric field would implement an XOR gate in a simple layered AF system. Finally, we investigated the critical current density, critical Néel phase and incubation time with various parameters.

## ACKNOWLEDGEMENT

This work was supported by the "GRI(GIST Research Institute)" Project through a grant provided by GIST in 2021, by National Research Foundation of Korea (NRF) funded by the

AUTHOR INFORMATION

T. H. Kim and S. Hwang contributed equally to this work.

APPENDIX A: A REORIENTATION IN THE COLLINEAR ANTIFERROMAGETS.

We consider the reorientation in the collinear AFM by choosing $D_y = 0$, $K_z = 2K_{z0}$ and including the Oersted field. For $K_z = K_{z0}$ $\varphi_{RE} > \pi/4$, a steady oscillation condition takes place before reorientation. Figure 6 shows the numerical and analytical results in the collinear AFM. In this system, exchange gain does not contribute to reorientation (see $\Delta E_{Ex}(t = 15 \text{ ps})$ in Fig. 6(a)). The reorientation is realized through the competition between two anisotropy energies, where $\Delta E_K(t = t_2)$ ($\Delta E_K(t = t_3)$) represents $\Delta E_{K,x}$ ($\Delta E_{K,z}$). Dark yellow indicates the analytical results, which are identical to the numerical results. The inset of Fig. 6 shows that $l_x$ is zero, implying that another way to achieve symmetric breaking is necessary. However, *l* is realigned along the *z-axis* at approximately 80 ps. Note that the reorientation direction is not affected by the bias field and is affected by the final phase of *l* just before reorientation.

**Figure captions**

Fig. 1. Schematic diagram of the antiferromagnetic device (a) and switching process (b-d) at $T<T_N$. The electric current flows in the $x$ direction, and the spin current is injected into the magnetic layer ($y$ direction) with polarization of the $z$-axis. Thus, the qAFM mode is excited where the Néel order $l$ oscillates on the $xy$ plane (b). In the critical Néel phase ($\varphi=\varphi_{RE}$), $l$ is reoriented along the z-axis ($\theta\sim0$) (c), and weak magnetization $m$ undergoes additional torque by SOT. As a result, $l$ is slightly tilted from the $x$-axis. Finally, after the current is turned off, $l$ is deterministically reversed. The green (yellow) color shows the initial (final) configuration.

Fig. 2. Antiferromagnetic switching triggered by SOT. The switching process is described in terms of the Néel order $l$ (a and c), magnetization $m$ (b and d) and magnetic energy (e). In Fig. (c) and Fig. (d), the arrows of $l$ and $m$ indicate the final phase. In (a), as the current density $J_c$ is applied, the Néel phase evolves up to $\varphi=\varphi_{RE}$ ($J_c=J_{RE}$). After incubation time $\tau_{in}$, $l$ is reoriented along the $z$-axis. The inset shows finite $l_x$ and $l_y$ in the reorientation state. $l_x$ is attributed to an additional torque between the SOT and $m$, where $m_z=l_xD_y/(2J)$. After $J_c$ is turned off, $l$ is reversed because of the biased $l_x\sim-0.026$. According to (e), Néel reorientation takes place when $\Delta E_T(t_2=45 \text{ ps}) > \Delta E_T(t_3=100 \text{ ps})$. (f) shows the magnitude of $l_x$ against the magnetic field along the *z-axis* in the reorientation state. The Oersted field in the *z* direction induces $m_z$ and, in turn, $l_x$.

Fig. 3. Antiferromagnetic switching and nonswitching depending on the sign of $D_y$ and $p_z$. The switching mode follows diode characteristics. For example, for $+D_y$, $+p$ ($-p$) triggers nonswitching (a) (switching (b)). Similarly, for $-D_y$, $+p$ ($-p$) triggers switching (a) (nonswitching (b)).

Fig. 4. The reorientation phase $\varphi_{RE}$ (a) and the reorientation current density $J_{RE}$ (b) as functions of $K_z$ and $D_y$ where the proportionality $n_1$ ($n_2$) indicates the ratio of variables $D_y$ ($K_z$) and $D_{y0}$ ($K_{z0}$). Incubation time $\tau_{in}$ (c) as a function of the damping constant $\beta$ and the current density $J_c$.

Fig. 5. Comparisons of SOT-driven switching between numerical (solid line) and analytical (dotted dark yellow line) results in the canted antiferromagnets.

Fig. 6. SOT-driven switching with respect to energetics (a) and dynamics (b) in a collinear AFM. In (b), the analytical results (solid line) are well matched to the numerical results (dotted dark yellow line).

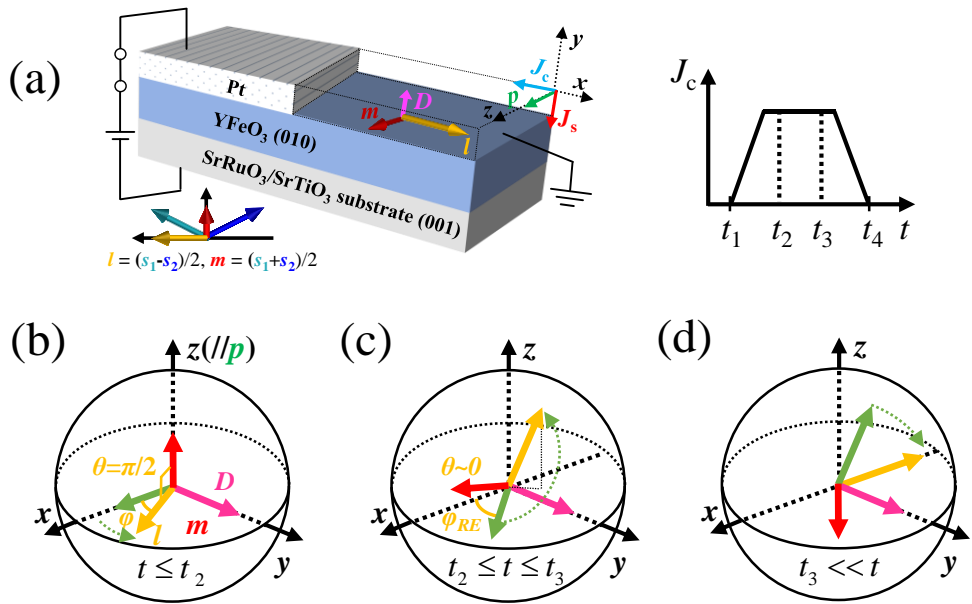

Figure 1

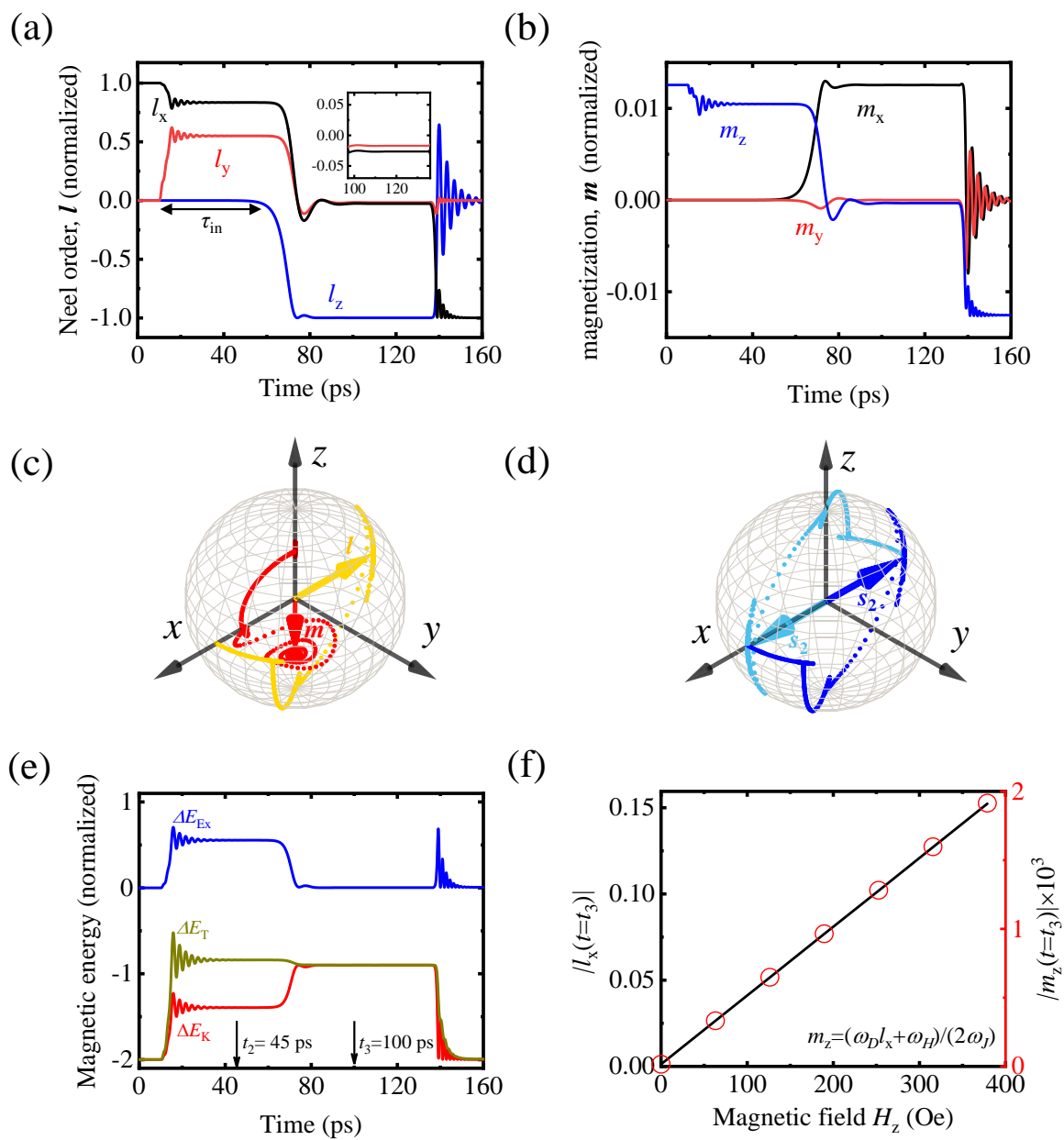

Figure 2

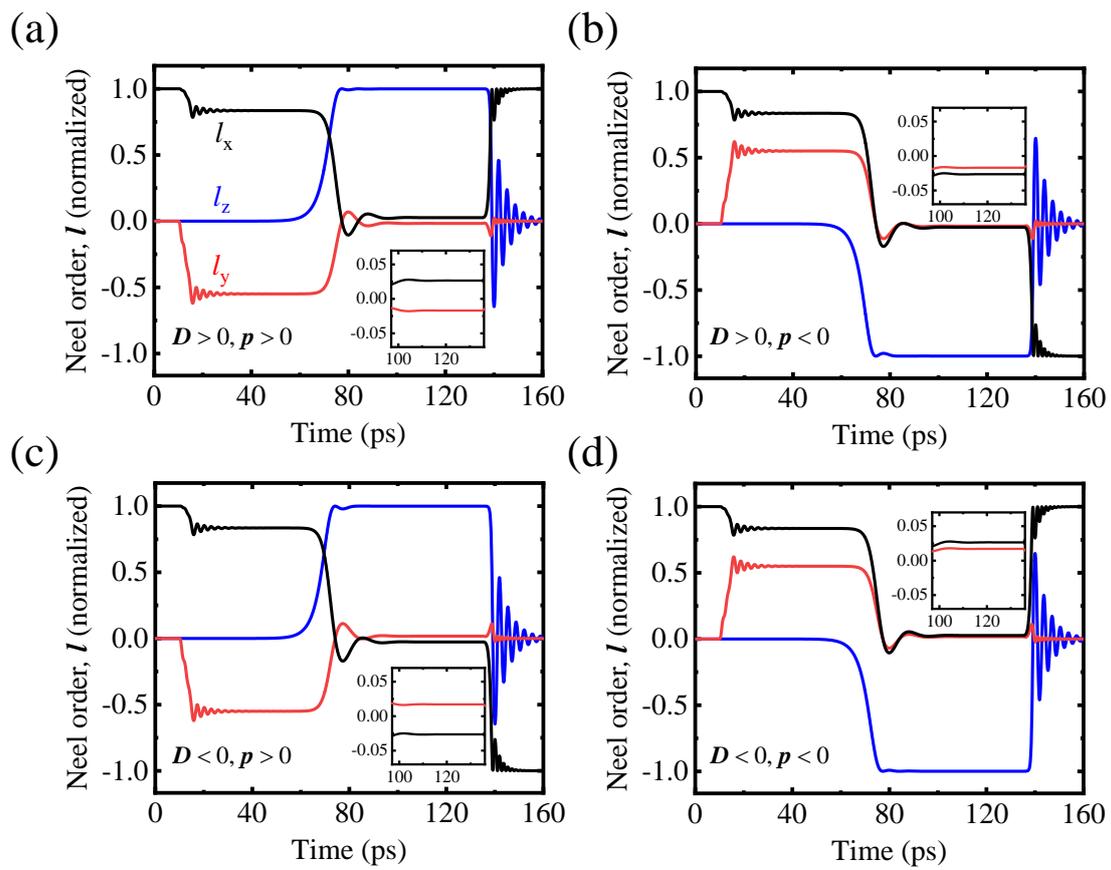

Figure 3

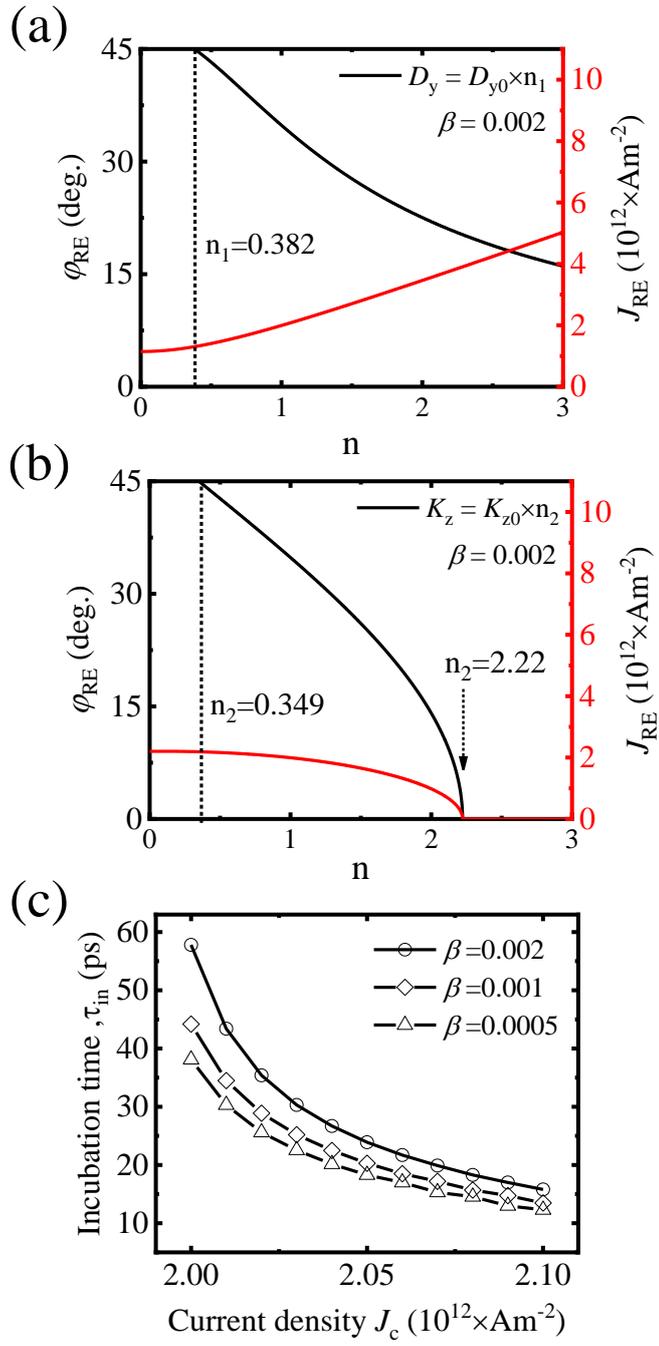

Figure 4

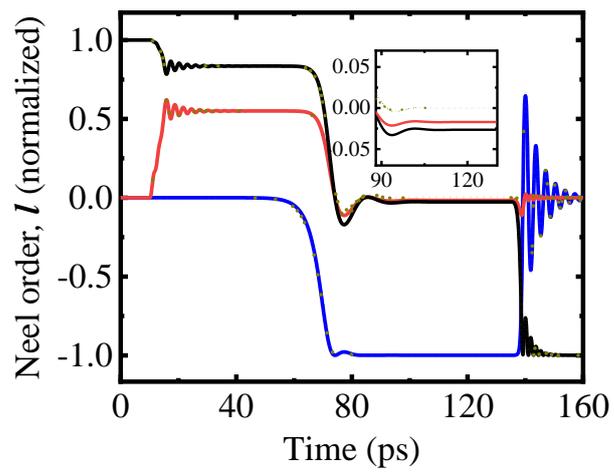

Figure 5

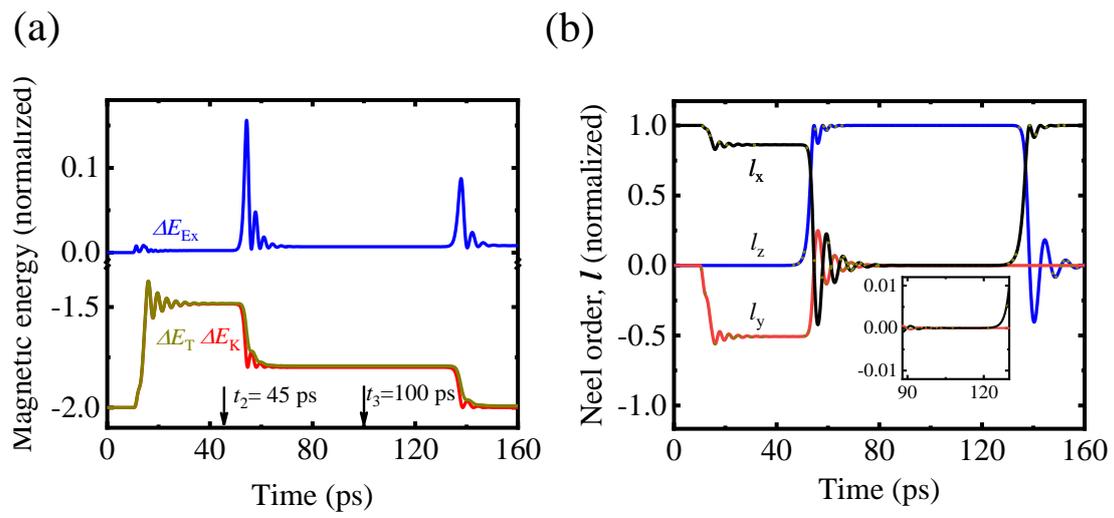

Figure 6
23